\documentclass[aps,prd,twocolumn,nofootinbib,longbibliography,10pt]{revtex4-2}
\usepackage{bm}
\usepackage{latexsym}
\usepackage{subfig}
\usepackage{amsmath,amsfonts,amssymb}
\usepackage{graphicx,epsfig}
\usepackage{psfrag}
\usepackage{amsthm}
\usepackage{bigints}
\usepackage{xcolor}
\usepackage{amsmath}
\usepackage[colorlinks=true,urlcolor=blue,citecolor=red,linkcolor=blue]{hyperref}
\usepackage{braket}
\usepackage[normalem]{ulem}
\usepackage{multirow}
\useunder{\uline}{\ul}{}
\usepackage{environ}
\usepackage{mathtools}
\usepackage{float}
\usepackage{relsize}
\usepackage{caption}
\usepackage{subcaption}

\newcommand{\hg}{\,_1{\rm F}_1}

\newcommand{\tra}{{\rm Tr_A}}
 
\def\o{\omega}

\def\k{\kappa}

\def\d{\delta}
\def\D{\Delta}

\def\t{\tau}
 
\def\p{\partial}
\def\f{\frac}

\newcommand{\be}{\begin{equation}}
\newcommand{\ee}{\end{equation}}
\newcommand{\bes}{\begin{equation*}}
\newcommand{\ees}{\end{equation*}}

\newcommand{\beq}{\begin{eqnarray}}
\newcommand{\eeq}{\end{eqnarray}}

\allowdisplaybreaks
\begin{document}
\title{\textbf{Derivative coupling in horizon brightened acceleration radiation: a quantum optics approach}}
\author{Ashmita Das}
\email{ashmita.phy@gmail.com}
\affiliation{Department of Physics, SRM University AP, Amaravati 522240, India}
\author{Anjana Krishnan}
\email{anjana.krishnan@srmap.edu.in}
\affiliation{Department of Physics, SRM University AP, Amaravati 522240, India}
\author{Soham Sen}
\email{sensohomhary@gmail.com, soham.sen@bose.res.in}
\affiliation{{Department of Astrophysics and High Energy Physics,}\\
{S.N. Bose National Centre for Basic Sciences,}\\
{JD Block, Sector III, Salt Lake, Kolkata 700106, India}}
\author{Sunandan Gangopadhyay}
\email{sunandan.gangopadhyay@gmail.com}
\affiliation{{Department of Astrophysics and High Energy Physics,}\\
{S.N. Bose National Centre for Basic Sciences,}\\
{JD Block, Sector III, Salt Lake, Kolkata 700106, India}}

\begin{abstract}
\noindent
Horizon Brightened Acceleration Radiation (HBAR) signifies a unique radiation process and provides a promising framework in exploring acceleration radiation in flat/ curved spacetime.
Its construction primarily relies on the transition probability of an atom falling through a high-Q cavity while interacting with a quantum field. The HBAR effect has typically been explored in the context of minimal coupling between the atom and the field amplitude. However, the minimally coupled models are affected by the infrared (IR) divergences that arise in the massless limit of the quantum fields in (1+1) dimensions. Thus, in the present manuscript,  we examine the HBAR process using both the point-like and finite size detectors coupled with the momentum of the field, which plays a crucial role in naturally resolving IR divergences. Our results suggest that the transition probability for the point-like detector is independent of its frequency. This can be interpreted as the influence of the local gravitational field which modifies the sensitivity of the detector to its frequency and broadens its effective frequency range. 
Through a comparative study based on the length of the detector, we find that for a detector with a smaller length, the steady state solution for the density matrix of the field vanishes. This may indicate the existence of a non equilibrium thermodynamic state under the condition of finite size detector-field interaction. 
These distinctive features are exclusive to the derivative coupling between the atom and the field, highlighting them as a compelling subject for future investigation.
\end{abstract}
\maketitle
\section{Introduction}\label{intro}
An elegant combination of individual disciplines such as general relativity, thermodynamics and quantum field theory (QFT) has brought to us many interesting phenomena. A direct consequence of this fusion is none other than the Hawking radiation from the black hole (BH) spacetime \cite{Hawking:1975vcx, Hawking:1976de}, cosmological particle production \cite{Unruh:1977ga, Gibbons:1977mu, Parker:1969au,Parker:1968mv, Parker:1971pt}, quantum entanglement \cite{Summers:1985pzz,Summers:1987ze, Mann:2005zza}, 
entanglement harvesting \cite{Reznik:2002fz,Pozas-Kerstjens:2015gta}, relativistic quantum information \cite{Ralph:2012mdp, Martin-Martinez:2011pqe} etc. Moving further the implementation of quantum optics in the studies of acceleration radiation in flat/ curved spacetime provided to us a parallel mechanism to explain the Unruh-Fulling (UF) effect \cite{Scully:2003zz}. Using a high-Q microwave cavity,  Scully et al. have shown in \cite{Scully:2003zz} that ground state atomic detectors when accelerated through the cavity, creates radiation where the transition probability of the detector can be increased to many orders in magnitude than that of the Unruh radiation \cite{Unruh:1976db, Birrell:1982ix, Crispino:2007eb}. Furthermore, in 2018 Fulling et al. extended this idea to the BH spacetime and presented a different kind of particle radiation called horizon brightened acceleration radiation (HBAR), unlike the Hawking radiation from the BH \cite{Scully:2017utk}. Consequently one obtains  the HBAR entropy which has a different genesis than the Bekenstein-Hawking entropy. 

The HBAR phenomenon is based on the virtual processes as we often witness in QFT  such as Lamb shift, Raman scattering, etc. 
Within the atom-field interaction the HBAR phenomenon can be outlined as follows. 
In the background of a BH spacetime, the two level atomic detectors are freely falling while they traverse through a cavity.  The cavity is positioned near the event horizon of the BH such as the Hawking radiated particles can be restricted to enter the cavity. To isolate a single field mode, a mode selector is installed which selects one cavity mode travelling in the opposite direction of the atoms so that 
the whole set up generates a relative acceleration between the atom and the field mode.
This relative acceleration incites the atom to drift away from its original position of virtual emission.  This triggers a non-zero probability of not absorbing the emitted virtual photon and  transforms a virtual photon into a real one in the final state of the system \cite{Scully:2003zz, Scully:2017utk}. Eventually the transition probability, temperature and the entropy associated with the HBAR can be estimated \cite{Scully:2003zz, Scully:2017utk}. HBAR scenario  has uncovered many notable insights which can be found in the following references \cite{Svidzinsky:2018jkp, Camblong:2020pme, Azizi:2020gff, Azizi:2021qcu, Azizi:2021yto, Sen:2022tru, Sen:2022cdx, Das:2022qpx, Das:2023rwg}. 

The interaction Hamiltonian for the HBAR model can be written as $\sim g \mu(\t)\Phi(x(\t))$, where $g, \,\mu(\t)$ depict coupling strength and the monopole moment of the atomic detector respectively. $\Phi(x(\t))$ is the massless scalar filed which interacts with the detector and $\t$ represents the proper time of the detector. This minimally coupled model resembles the Hamiltonian of the Unruh-DeWitt (UD) detector interaction with a massless scalar field in the standard UF effect \cite{Unruh:1976db, Birrell:1982ix}.

Identifying this similarity, we turn our focus towards the well known infrared (IR) divergence problem of the standard UF effect for a massless scalar field in $(1+1)$ dimensions. This problem is inbuilt in the Wightman function of a minimally coupled massless field in $(1+1)$ dimensions \cite{Juarez-Aubry:2021tae, Bunney:2023vyj,Takagi:1986kn}. This IR divergence can be tackled by using a cut-off scale, however, final density matrix and transition probability of the detectors become dependent on the chosen cut-off scale \cite{Pozas-Kerstjens:2015gta,Martin-Martinez:2014qda, Juarez-Aubry:2014jba}. Subsequently, it was shown that this IR ambiguity can naturally be removed by considering the detector linearly coupled to the proper time derivative of the field amplitude \cite{Juarez-Aubry:2014jba, Teixido-Bonfill:2024phf,Juarez-Aubry:2021tae,Bunney:2023vyj,Martin-Martinez:2014qda, Raval:1995mb, Wang:2013lex, Tjoa:2020eqh, Louko:2014aba, Juarez-Aubry:2018ofz}. Furthermore the short distance behaviour of derivative coupling in $(1+1)$ dimensions mimics the $(3+1)$ dimensional minimal coupling \cite{Juarez-Aubry:2014jba, Juarez-Aubry:2018ofz}, which leads to the dualities between the derivative and minimally coupled models \cite{Perche:2023vhd}. Thus, particle detector model with derivative coupling possesses considerable importance in controlling the IR divergence in QFT and offers a simple $(1+1)$ dimensional framework that captures the essential features of minimally coupled particle detector models in $(3+1)$ dimensions, so that complicated treatments of $(3+1)$ dimensions can be avoided. In recent times, several attempts have been taken to study the implications of derivative coupling such as derivative coupling supports the entanglement harvesting when the particle detectors are in causal contact in flat spacetime \cite{Teixido-Bonfill:2024phf}, construction of relativistic communication channel between two localised detectors in arbitrary curved spacetime \cite{Kasprzak:2024rzj}, establishing dualities between minimally coupled particle detector model and derivative coupled model in the limit of large energy gaps \cite{Perche:2023vhd}, etc. Furthermore, using the derivative coupling approach, a UD detector executing uniform circular motion is considered in a $(2+1)$ dimensional background, with the field initially in a thermal state \cite{Bunney:2023vyj}. Obtaining the effective temperature associated with the detector, it was shown in \cite{Bunney:2023vyj} that the effect due to the acceleration radiation can indeed be detected. 
Under certain conditions, the existence of cooling Unruh effect has also been reported in \cite{Bunney:2023vyj}, making the model to be experimentally promising. 
It is also observed in \cite{Pozas-Kerstjens:2016rsh, Lopp:2020qwx} that the interaction of the atom and derivative of the scalar field captures certain characteristics of the interaction between the atom and electromagnetic field. 
Derivative coupled models can also be experimentally tested by employing superconducting qubit as a particle detector, coupled to a transmission line which mimics a $(1+1)$ dimensional massless field \cite{McKay:2017dte, Janzen:2022mws}. 

Motivated by the implications of derivative coupling, we explore a spontaneous question, what would be the fate of HBAR phenomenon when the atomic detector and the field exhibit derivative coupling? Specifically, we consider a high Q microwave cavity set up in the vicinity of the Schwarzschild BH as described in \cite{Scully:2017utk}. The stream of two level atomic detectors are falling freely from infinity towards the BH through the cavity set up. Within the cavity one or few modes of the quantum matter field can be confined and a mode selector chooses one cavity mode,  enabling a relative acceleration between the field mode and the infalling atoms. In this setup, we consider atoms coupled to the momentum of a massless scalar photon, taking into account both point-like and finite size descriptions of the atoms. 
We summarize our findings below. 
  \begin{enumerate}
  \item
We compute the transition probability of the system up to the first order of perturbation theory for both the point-like and finite size detectors. For the point-like detector, we obtain an enhanced transition probability than as reported in \cite{Scully:2017utk}. On the other hand the transition probability for the finite size detector turns out to be dependent on the size of the detector and exhibits suppression for larger extent of the detector. 
  \item 
For point-like case, we find the temperature of the thermal bath due to the HBAR to be $T=\f{\hbar c^3}{8 \pi G M_{\rm BH}k_B}$, where $\hbar$, $c$, and $k_B$ represent reduced Planck's constant, speed of light in vacuum and the Boltzmann constant respectively. $G$ is the Newton's gravitational constant and $M_{\rm BH}$ depict the mass of the Schwarzschild BH. 
For the finite size detector two conditions emerge such as $L\o\gtrsim 1$ and $L\o\lesssim 1$, where $L$ represents the length of the detector and $\o$ stands for the frequency of the same. Physically we interpret these conditions as larger and smaller length detectors for a fixed frequency $\o$. Thus, for $L\o\gtrsim 1$, we obtain a well defined temperature of the thermal bath, similar to the point-like case whereas for $L\o\lesssim 1$, the transition probability displays non-Planckian characteristics, indicating that no well defined temperature can be assigned.

\item
 We find the microscopic change in the density matrix of the field due to the radiation and subsequently derive the steady state solution for the same. In case of the point-like detector, the steady state solution remains same as in \cite{Scully:2017utk}. For the finite size detector, under the condition $L\o\gtrsim 1$, the steady solution is obtained, however, it becomes zero for $L\o\lesssim 1$.
 \item 
In this framework, the entropy associated to HBAR  exhibits a similar area entropy relation as one obtains for the standard BH solutions and in \cite{Scully:2017utk}. 
\item
We implement the phenomenon of Wien displacement and make a comparative study between the wavelengths of the emitted radiation, which correspond to the maximum transition probability in both the minimal and derivative coupled models. In case of point-like and finite size detector with the condition $L\o\gtrsim 1$, the nature of the $\lambda$ vs. temperature plot exhibit the same pattern. 
 \end{enumerate}
The inclusion of derivative coupling between the atom and the field leads to modified behaviours that not only resolve the IR divergence problem found in the minimally coupled model but also introduce unique characteristics in the HBAR phenomenon. 

\noindent The present manuscript is organised as follows. In the upcoming section (\ref{trajectory_atom}), we describe the trajectory of the atomic detector and write the field equations in the background of Schwarzschild spacetime. Sections (\ref{pointlike}) and (\ref{finite_size}) discuss the interaction of point-like and finite size detectors with the momentum of a massless scalar photon and derive the corresponding excitation probabilities, the density matrix of the field, steady state solution and area-entropy law. 
 In section (\ref{wien_1}), relating to the Wien's displacement law, we examine how the radiation wavelength varies with respect to the temperature of the bath and present a comparative study between the amplitude and derivative coupled models. 
\section{Trajectory of the atomic detector and mode solution for the field in the background of Schwarzschild spacetime}\label{trajectory_atom}
In this section, we review the trajectory of a freely falling atomic detector in the background of a Schwarzschild BH where we consider the $t-r$ part of the metric as follows \cite{Scully:2017utk}, 
\beq
ds^2=\left(1-\f{r_g}{r}\right)- \left(1-\f{r_g}{r}\right)^{-1} dr^2, 
\label{sch_1}
\eeq
where $r_g=\f{2GM_{BH}}{c^2}$ is the radius of the event horizon of the BH. At this stage we introduce a dimensionless coordinate where $r_g,\,r_g/c$ are treated as the unit of the distance and time respectively. Therefore, following \cite{Scully:2017utk} we write.
\beq
r\to r_g r,  \quad \quad t\to \left(r_g/c\right)t,  \quad \quad \o\to \left(c/r_g\right)~.
\eeq
Note that in the transformation the parameters at the right hand side of the arrow head represent dimensionless parameters. 
In terms of dimensionless Schwarzschild coordinates, the trajectory of the infalling atom with respect to its dimensionless proper time $\t$ becomes, 
\beq
\f{dr}{d\t}=-\f{1}{\sqrt{r}},\quad\quad \f{dt}{d\t}=\f{r}{r-1}.
\label{traj_1}
\eeq
Solving these equations one obtains, 
\beq
&\t=-\f{2}{3} r^{3/2}+{\rm const.}\label{tau_r_1}\\
&t=-\f{2}{3} r^{3/2} - 2\sqrt{r}- {\rm ln}\left(\f{\sqrt{r}-1}{\sqrt{r}+1}\right)+{\rm const.}
\label{t_r_1}
\eeq
At this stage, we follow the procedure developed in \cite{Scully:2003zz,Scully:2017utk} and consider the Klein Gordon equation for a massless scalar photon with wave function $\phi$ as, 
\beq
\f{1}{\sqrt{-g}}\p_{\mu}\left(\sqrt{-g}\,g^{\mu\nu} \p_{\nu}\right)\phi = 0~.
\eeq
Imposing the s-wave approximation and a coordinate transformation $r_{*}=r+{\rm ln}(r-1)$, in the above equation one obtains, 
\beq
\left(\f{\p^2}{\p t^2}- \f{\p^2}{\p r_{*}^{2}} \right)\phi =0
\label{scalar_eqn_1}
\eeq
The solution of the Eq. (\ref{scalar_eqn_1}) turns out to be, 
\beq
\phi=e^{i \nu (t-r_{*})}=e^{i \nu\left[t-r-{\rm ln}(r-1)\right]}, 
\label{scalar_sol_1}
\eeq
where, $\nu$ depicts the frequency of the photon field. 
In the upcoming section we turn our focus to explore the HBAR phenomenon in case of the point-like detector interacting with the momentum of a massless scalar photon. 
\section{HBAR for the point-like detector coupled to the momentum of the field in the background of Schwarzschild BH}\label{pointlike}
We write the interaction Hamiltonian of a point-like detector with the momentum of a massless scalar photon as follows \cite{Crispino:2007eb}, 
\beq
  H_{int}= \hbar\,\mathtt{g}\,\mu(\tau) \sqrt{-g} \, g^{00} \, \partial_0\left(a_{\nu}e^{-i \nu t(\t)+i \nu r_{*}(\t)}+{\rm H.C}\right)\nonumber\\
  \label{hamiltonian_1}
 \eeq
 where $\mathtt{g}$ is the coupling constant and $\mu(\tau)=\sigma e^{-i\omega\tau}+\sigma^{\dagger}e^{i\omega\tau}$ is the monopole moment operator of the detector. The momentum of the field in the curved spacetime can be defined as, 
 $\pi (\t)=\sqrt{-g} \, g^{00} \, \left[\partial_0\,\Phi(\t)\right]$ \cite{Crispino:2007eb}. Here $\Phi (\t)=\left(a_{\nu}e^{-i \nu t(\t)+i \nu r_{*}(\t)}+{\rm H.C}\right)$, represents the massless scalar field. 
$g_{00}$ denotes the zeroth component of the $(3+1)$ dimensional Schwarzschild metric and $\sqrt{-g}$ is the determinant of the metric tensor $g_{\mu\nu}$. In the dimensionless Schwarzschild coordinates $g^{00}$  becomes $\frac{r}{r-1}$.
$\sqrt{-g} = r^2 \sin\theta$, which simplifies to $\sqrt{-g} = r^2$ when restricted to the equatorial plane, {\it i.e.}, at $\theta = \frac{\pi}{2}$. 
The initial and the final state of the system can be represented as, $\ket{0, a}$ and $\ket{1_{\nu}, b}$ respectively, where $(0,\,1_{\nu})$ are the vacuum and 1-particle state of the matter field with a particular frequency $\nu$ and  $(a,\,b)$ depict the ground and excited state of the detector. At this stage we refer our readers to \cite{Scully:2017utk} for an elaborate review of this set up. Therefore, decomposing the field operator and operating the time derivative on the field modes we obtain, 
\begin{align}
    H_{int}&= i\,\hbar\,\mathtt{g}\nu\,\mu(\tau)\frac{r^3}{r-1}\left(a_{\nu}^{\dagger}e^{i\nu(t-r_*)}-a_{\nu}e^{-i\nu(t-r_*)}\right)~.
     \label{hamiltonian_2}
\end{align}
Using Eqs. (\ref{tau_r_1}), (\ref{t_r_1}) and $r_{*}=r_{*}(r)$, the transition probability is found to be, 
\begin{align}
 &P_{exc}= \frac{1}{\hbar^2}\bigg|\int d\tau \bra{1,b}H_{int}\ket{0,a}\bigg|^2 = \mathtt{g}^2\nu^2\nonumber\\
& \bigg|\int_1^{r'}dr \left(\frac{r^{7/2}}{r-1}\right)
 e^{-i\nu(\frac{2}{3}r^{3/2,}+r+2\ln(r^{1/2}-1)+2r^{1/2})}e^{\frac{-2i\omega}{3}r^{3/2}}\bigg|^2
 \label{point_exc_1}
    \end{align}
    where $r'$ is considered to be $1\lesssim r'< 2$, depicting the near horizon region of the BH. Note from the previous discussion that, in the dimensionless Schwarzschild coordinates, the radius of the event horizon turns out to be $1$. 
Substituting $y= r^{3/2}$ in Eq. (\ref{point_exc_1}), we find, 
  \beq
 && P_{exc}=\frac{4\mathtt{g}^2\nu^2}{9}\bigg|\int_{1}^{y'}dy\left(\frac{y^2}{y^{2/3}-1}\right)\nonumber\\
  &&e^{-i\nu(\frac{2}{3}y+y^{2/3}+2y^{1/3}+2\ln(y^{1/3}-1))}e^{-\frac{2i\omega}{3}y}\bigg|^2~,
  \label{point_exc_2}
  \eeq
where $y'=r'^{\frac{3}{2}}$ is the upper limit of integration. Since we are still working in the near horizon regime, we restrict $y'$ to the range $1\lesssim y'<2$. 
Substituting $x=\frac{2\omega}{3}(y-1)$ one obtains $x<\frac{2\omega}{3}$ and thus the above equation can be recast as follows, 
\begin{align}
P_{exc}= \frac{\mathtt{g}^2\nu^2}{\omega^2}\bigg|\int_{0}^{x'} dx \left(\frac{\omega}{x}+3\right)e^{-ix\left(1+\frac{2\nu}{\omega}\right)}x^{-2i\nu}\bigg|^2~.
  \label{point_exc_3}
    \end{align}
 where $x'=\frac{2\omega}{3}(y'-1)$.  It is important to note that $x<\frac{2\omega}{3}\implies \frac{x}{\omega}<\frac{2}{3}<1$. Some important observations are now in order. The functions $x^{-2i\nu}$ and $e^{-ix \left(1+\frac{2\nu}{\omega}\right)}$ are both oscillatory in nature.
 \begin{figure}[ht!]
 \begin{center}
 \includegraphics[scale=0.21]{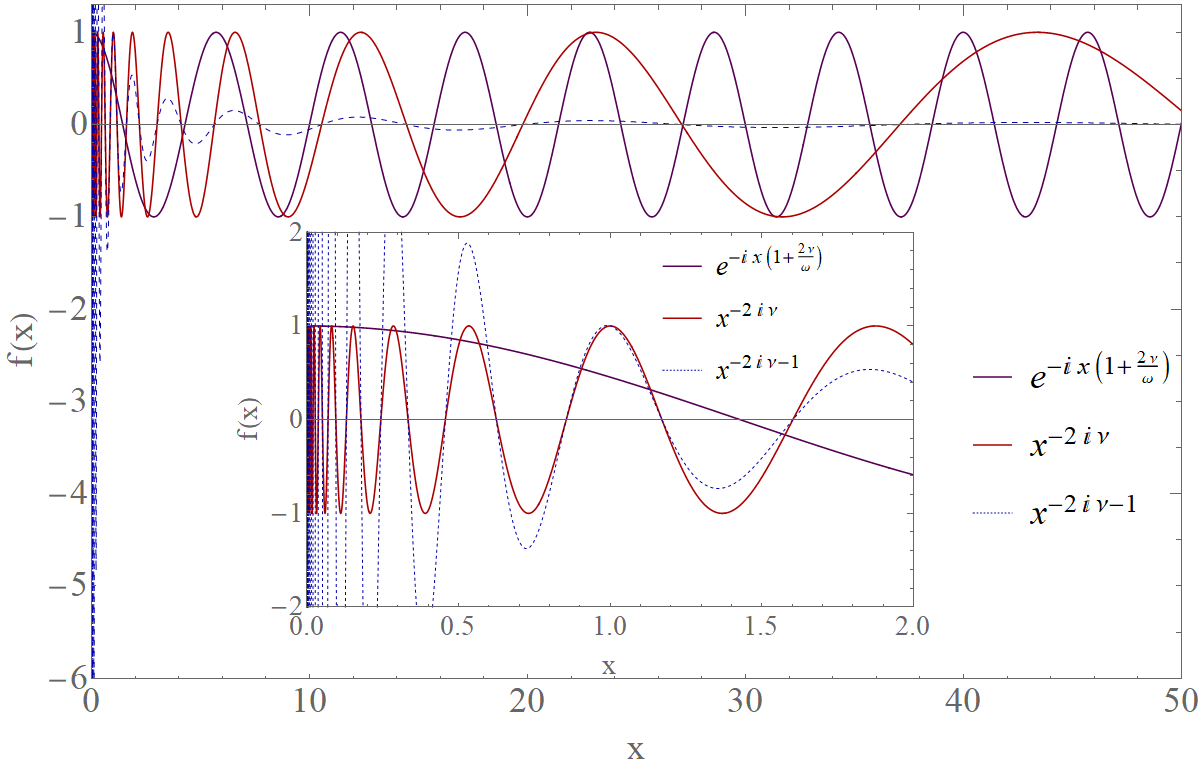}
 \caption{Real part of the functions $f(x)=\{e^{-ix\left(1+\frac{2\nu}{\omega}\right)},~x^{-2i\nu},~x^{-2i\nu-1}\}$ are plotted against $x$ for $\frac{\nu}{\omega}=\frac{1}{20}$.\label{Russian_Doll}}
 \end{center}
 \end{figure}
 In Fig.(\ref{Russian_Doll}), we plot the real part of the following functions $f(x)=\{e^{-ix\left(1+\frac{2\nu}{\omega}\right)},~x^{-2i\nu},~x^{-2i\nu-1}\}$ with respect to $x$, while keeping $\nu=5$ and $\omega=100$ fixed in the dimensionless unit. From Fig.(\ref{Russian_Doll}), it can be perceived that in the large $x$ limit $x^{-2 i \nu}$ varies slowly with respect to the oscillatory $e^{-i x \left(1+\frac{2\nu}{\omega}\right)}$ function and the effect of the $x^{-1-2i \nu}$ almost dies out. Conversely, we observe from the inset plot that, for small $x$ regime the exponential function is slowly varying in comparison to $x^{-2i\nu}$ as well as $x^{-2i\nu-1}$. This functional behaviour of $x^{-2i\nu}$ is also known as the ``Russian doll behaviour" \cite{Camblong:2020pme}. As a result, the dominant contribution to the integral in eq.(\ref{point_exc_3}), shall come from the regime where $0\leq \frac{x}{\omega}\lesssim 1$. On the other hand, in large $x$ regime, due to the highly oscillatory nature of the exponential function compared to the $x^{-2i\nu}$, the overall contributions to the integral can be neglected. As a result, it is possible to extend the upper limit of integration from $x'$ to $\infty$ without affecting the analysis. This recasts Eq.(\ref{point_exc_3}) in the following form, 
\begin{equation}\label{point_exc_3b}
\begin{split}
  P_{exc}= \frac{\mathtt{g}^2\nu^2}{\omega^2}\bigg|\bigintsss_{0}^{\infty} dx \left(\frac{\omega}{x}+3\right)e^{-ix\left(1+\frac{2\nu}{\omega}\right)}x^{-2i\nu}\bigg|^2~.
\end{split} 
\end{equation}
Using $u= x\left(1+\frac{2\nu}{\omega}\right)$ and performing the integral, the transition probability turns out to be, 
\begin{align}     
P_{exc}&=\frac{\mathtt{g}^2\nu^2}{\omega^2}\bigg|\Gamma(-2i\nu)e^{-\pi\nu}(\omega-4\nu)\bigg|^2\nonumber\\
&=\frac{g^2\nu\pi}{\left(1+\frac{2\nu}{\omega}\right)^2}\left(1-\frac{4\nu}{\omega}\right)^2\frac{1}{e^{4\pi\nu}-1}~, 
\label{point_exc_4}
\end{align}
where we use the identity $\Gamma(ix)= \frac{\pi}{x\sinh(\pi x)}$. It is apparent that for $\nu \gg 1$, the transition probability is exponentially suppressed. For characteristic atomic frequencies, $\o \gg 1$, therefore, it is reasonable to focus on the domain where, in general $\nu \ll \o$. Under this condition the transition probability emerges as, 
\begin{align}
    P_{exc}&= \mathtt{g}^2\nu\left(\frac{1}{e^{4\pi\nu}-1}\right)~.
    \label{point_exc_5}
\end{align}
In dimensionful units, Eq. (\ref{point_exc_5}) can be recast as, 
\begin{align}
    P_{exc}= \frac{\mathtt{g}^2\nu r_g\pi}{c}\frac{1}{e^{4\pi\nu r_g /c}-1}~.
    \label{point_exc_6}
\end{align}
Subsequently, we obtain the absorption probability by altering $\nu\to -\nu$, which yields the absorption probability ($P_{abs}$) as below,
\be
 P_{abs}= e^{4\pi\nu r_g/c}P_{exc}~.
 \label{point_abs_1}
\ee
Eq. (\ref{point_exc_6}) demonstrates that the transition probability of the system under the leading order approximation of the time dependent perturbation theory, becomes independent of the detector's frequency. We propose that momentum coupling in curved spacetime explicitly introduces a connection between the spacetime curvature, {\it i.e.,} the gravitational field, and the detector. This interaction may provide enough energy to the detector to assist its transition without the need for exact resonance at a particular frequency. Consequently, the resulting transition probability becomes independent of the detector's frequency \cite{Juarez-Aubry:2014jba}. 
We illustrate how the transition probability varies with the field frequency and provide a comparative analysis of $P_{exc}$ as derived from both the minimally coupled model \cite{Scully:2017utk} and the momentum coupled model. For the purpose of the plotting, we treat all the parameters in dimensionless unit. 
 \begin{figure}[h!]
    \centering
    \includegraphics[width=\linewidth]{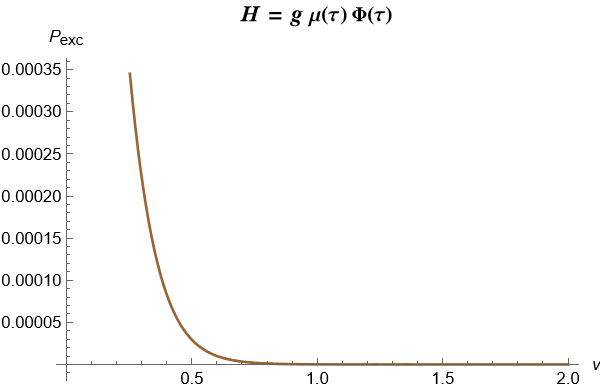} 
    \caption{$P_{exc}$ vs $\nu$ plot for minimal coupling between the detector and the field. $\mathtt{g}$ is set to be 1 while $\o=20$}
    \label{fulling_trans}
\end{figure}
 \begin{figure}[h!]
    \centering
    \includegraphics[scale=0.75]{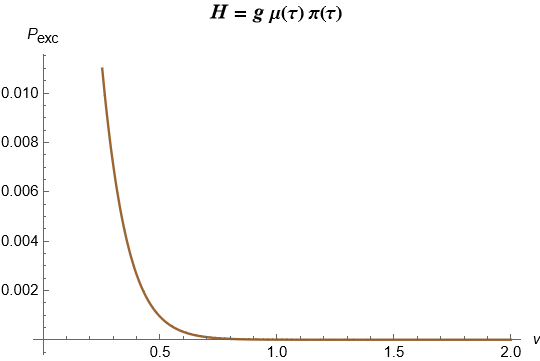}
    \caption{ $P_{exc}$ vs $\nu$ plot for the detector coupled with the momentum of the field. $\mathtt{g}$ is set to be $1$}
    \label{derv_trans_point}
\end{figure}
\subsection{Density matrix for the field mode and HBAR entropy: point-like case}\label{point_density}
\noindent In this section, we find the entropy flux corresponding to the HBAR in case of the point-like detector interacting with the momentum of the field. Following the treatment from quantum optics as used in \cite{Scully:2003zz,Scully:2017utk,Azizi:2021qcu}, we first write the microscopic change in the field density matrix as, $\d \rho_{i}$ due to a single atom-field interaction. Consequently, the macroscopic change in the same due to the $\D \mathcal{N}$ number of atoms can be written as \cite{Scully:2003zz,Scully:2017utk,Azizi:2021qcu}, 
\beq
&&\Delta \rho=\, \sum_{i} \d \rho_i=\,\D \mathcal{N}\,\d \rho=\,\k\,\D t\,\d \rho
\label{den_fld_1}\\
&&\implies\, \f{\D \rho}{\D t}=\,\k\,\d \rho~.
\label{den_fld_2}
\eeq
Here, $\k$ denotes the rate at which the atoms fall into the event horizon of the BH. Using the Lindblad master equation for the density matrix one obtains, 
\beq
\dot{\rho}_{n,n}= &-& R_a \left[n \rho_{n,n} - (n+1) \rho_{n+1, n+1}\right] \nonumber\\
&-& R_e \left[(n+1) \rho_{n,n} - n \rho_{n-1, n-1}\right], 
\label{rho_lin}
\eeq
where, $R_e,\,R_a$ symbolize emission and absorption rates of the photons in the cavity by the atom and these rates are defined as, $R_{e/a}= \k\,P_{\rm exc/abs}$. 
Thus, in terms of the excitation and absorption rates, the steady state solution becomes, 
\beq
\rho_{n,n}^{\mathcal{S}}=\, \left(\f{R_e}{R_a}\right)^{n}\,\left( 1- \f{R_e}{R_a}\right).
\label{rho_ss_1}
\eeq
Using Eqs.(\ref{point_exc_6}) and (\ref{point_abs_1}), the steady state solution for the density matrix of the field turns out to be, 
\begin{align}
    \rho_{n,n}^{\mathcal{S}}&= \left(\frac{P_{exc}}{P_{abs}}\right)^n\left(1-\frac{P_{exc}}{P_{abs}}\right)\notag\\
    &= e^{-4\pi\nu nr_g/c}(1-e^{-4\pi\nu r_g/c})~.
    \label{density_point_1}
\end{align}
Furthermore, using the density matrix of the field we aim to find the Von Neumann entropy for the system. Thus, the time rate of change of entropy within the cavity due to the photon emission becomes,  
\beq
\dot{S}_p=\,-\,k_{B}\,\sum_{n,\nu}\,\dot{\rho}_{n,n}\,{\rm ln}(\rho_{n,n}).
\label{hbar_entropy_1}
\eeq
Using the steady state solution of the density matrix, the above equation can approximately be written as, 
\beq
\dot{S}_p\,&\approx&\,-\,k_{B}\,\sum_{n,\nu}\,\dot{\rho}_{n,n}\,{\rm ln}(\rho^{\mathcal{S}}_{n,n}).
\label{hbar_entropy_2}
\eeq
Subsequently, using Eq.  (\ref{density_point_1}) in Eq.  (\ref{hbar_entropy_1}) we obtain, 
\beq
\dot{S}_p&=& -k_B\sum_{n,\nu}\dot{\rho}_{n, n}\bigg[\frac{-4\pi\nu nr_g}{c}+ \ln(1-e^{-4\pi\nu r_g/c})\bigg] \notag \\   
    &\approx& k_B \sum_{n,\nu}\dot{\rho}_{n, n}\frac{4\pi\nu r_g n}{c}= k_B \frac{4\pi r_g}{c}\sum_{\nu} \dot{\bar{\eta}}_{\nu} \,\nu ~.
     \label{eq:entropy approximation}
\eeq
Here, $\dot{\bar{\eta}}_{\nu}=\sum_n n \dot{\rho}_{n,n}$ depicts flux of the produced photons in the cavity. 

\noindent
The area of the BH can be written as $A_{\rm BH}=\,4 \pi\,r_{g}^{2}=\f{16 \pi G^2\,M^{2}_{\rm BH}}{c^4}$, and we write the rate of change of the rest mass of the BH as below,
\beq
\dot{M}_{\rm BH}=\,\dot{M}_{\rm photon}+\,\dot{M}_{\rm atom}~,
\label{mass_change_time_1}
\eeq
where $\dot{M}_{\rm atom}$ represents the rate of change of mass of the BH due to the injection of the atomic cloud, and $\dot{M}_{\rm photon}$ depicts rate of change of mass of the BH  due to the extraction of energy by the emitted photons from the rest mass energy of the BH. 
Differentiating the area of the BH with respect to time we obtain, 
\be
\dot{A}_{\rm BH}= \frac{32\pi G^2 M_{\rm BH}\dot{M}_{\rm BH}}{c^4}. 
\label{area_bh_1}
\ee
Using the above equation and area of BH we write, 
\beq
\dot{A}_{\rm BH}=\f{2 \,\dot{M}_{\rm BH}A_{\rm BH}}{M_{\rm BH}}=\f{2\, A_{\rm BH}\,(\dot{M}_{\rm photon}+\,\dot{M}_{\rm atom})}{M_{\rm BH}}~.
\label{area_ratio_1}
\eeq
Eq. (\ref{area_ratio_1}) can also be expressed as follows,  
\beq
\dot{A}_{\rm BH}= \dot{A}_{\rm photon}+\,\dot{A}_{\rm atom}~,
\label{area_change_4}
\eeq
which, upon comparison with Eq. (\ref{area_ratio_1}), implies 
\beq
\dot{A}_{\rm photon (atom)}&=&\f{2\,A_{\rm BH} \dot{M}_{\rm photon(atom)}}{M_{\rm BH}}\nonumber\\
&=&\f{32 \pi G^2\,M_{\rm BH} \dot{M}_{\rm photon(atom)}}{c^4}~.
\label{area_mass_1}
\eeq
Eq.  (\ref{area_change_4}) indicates that the rate of change of the area of the BH is a summation of the rate of change of the area due to the photon emission and injection of atomic cloud near the BH. 
At this stage, we focus on the rate of change of entropy of the BH, only due to the photon emission and write, 
\beq
\dot{S}_p=k_B \frac{4\pi r_g}{c \hbar}\sum_{\nu} \hbar \,\dot{\bar{\eta}}_{\nu} \,\nu = k_B \frac{4\pi r_g}{c \hbar} \,\dot{M}_{\rm photon} c^2~, 
     \label{entropy_approx_1}
\eeq
where, using Eq. (\ref{area_mass_1}) and $r_g=\f{2 G M_{\rm BH}}{c^2}$, we arrive at the following expression, 
\beq
 \dot{S}_p=  \frac{k_B c^3}{4\hbar G}\dot{A}_{\rm photon}~.
 \label{final_entropy_1}
\eeq
Eq.  (\ref{final_entropy_1}) demonstrates the relation between the rate of change of HBAR entropy due to the emission of the photon and the area of the BH. For the detailed derivation of the density matrix we refer our readers to the Appendix (\ref{subsec:density matrix}). 
\section{HBAR for the finite size detector coupled to the momentum of the field in the background of Schwarzschild BH}\label{finite_size}
\begin{figure}[h!]
 \centering
  \includegraphics[scale=0.16]{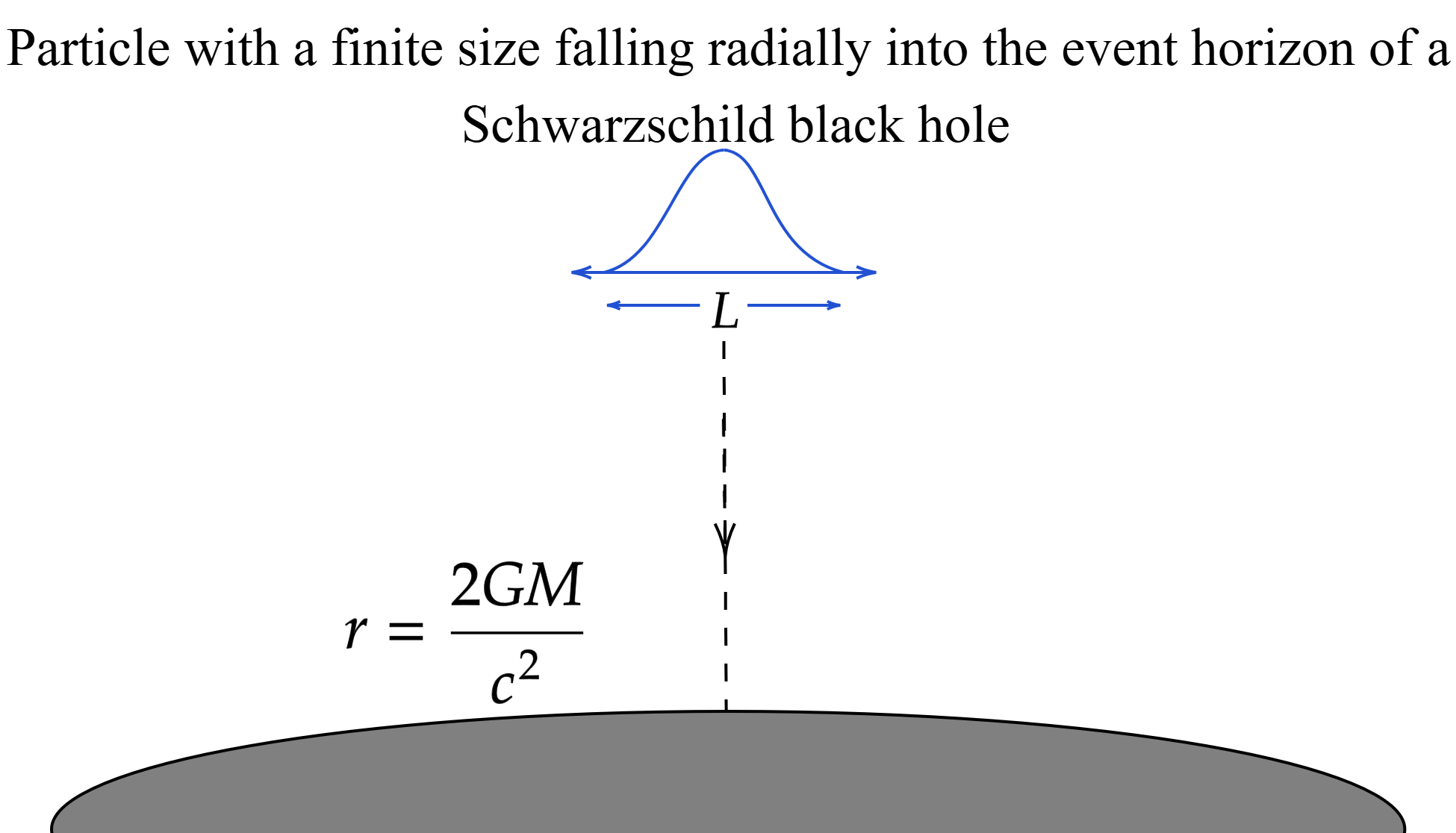}
   \caption{ }
    \label{smearing_picture}
\end{figure}

In this section, we consider the interaction where a finite size detector is coupled to the momentum of the scalar photon in the Schwarzschild BH spacetime. We explore the implications of finite size detector, as in reality, any kind of detectors possess spatial extension. In addition, a  point-like detector interacting with fields suffers from the ultraviolet (UV) divergences due to the singularity that appear in the field correlation functions at coincidence limit \cite{Schlicht:2003iy, Langlois:2005nf, Louko:2006zv, Louko:2006yf, Martin-Martinez:2012ysv}. To resolve the UV ambiguity in case of the point-like detector, several regularization schemes are introduced. For the regularisation schemes, we refer our readers to \cite{Schlicht:2003iy} and the references therein. 
We write the interaction Hamiltonian for the finite size detector as follows, 
\beq
H_{int}&=& \hbar\,\mathtt{g} \,\left(\sigma e^{-i\omega\tau}+\sigma^{\dagger}e^{i\omega\tau}\right)\int dr'F(r') \sqrt{-g}  g^{00} \partial_0\phi \notag\\
&=& i\,\hbar\,\mathtt{g}\,\nu\,\left(\sigma e^{-i\omega\tau}+\sigma^{\dagger}e^{i\omega\tau}\right)\int dr'F(r')\frac{r^3}{r-1}\nonumber\\
&& \left(a_{\nu}^{\dagger}e^{i\nu(t-r_*)}-a_{\nu}e^{-i\nu(t-r_*)}\right)
 \label{finite_hamil_1}
\eeq
Here $\mathtt{g}$ represents the coupling strength of the finite size detector with the field and $F(r)$ is the smearing function of the detector having a gaussian structure given  by \cite{Martin-Martinez:2012ysv}, 
\begin{equation}\label{F(r)}
F(r) = \frac{\exp[-(r-1)^2/2L^2]}{L\sqrt{2\pi}}~.
\end{equation} 
In the figure (\ref{smearing_picture}), we present a schematic ,illustrating a finite size detector falling radially towards the Schwarzschild BH, maintaining the horizontal orientation. For simplicity, we consider the detector to be oriented horizontally, so that complications arising from changes in the detector's proper time due to the gravitational effects of the BH can be avoided. 
The finite size detector is coupled with the momentum of a massless scalar photon. The length of the detector is so small that we have ignored any kind of tidal effects in the detector arm.

Following the similar approach as followed in case of the point-like detector, we write, 
\beq
     &&P_{exc}= \frac{16g^2\nu^2}{162\pi L^2}\bigg|\int_1^{\infty}dy e^{-\frac{2i\omega}{3}y}\int_1^y dy' 
   \left(\frac{y'^{5/3}}{y'^{2/3}-1}\right)\nonumber\\
   &&e^{-\frac{(y'^{2/3}-1)^2}{2L^2}}e^{-i\nu(\frac{2}{3}y'+y'^{2/3}+2y'^{1/3}+2\ln(y'^{1/3}-1))}\bigg|^2
\eeq
where, we substitute $y= r^{3/2}$. Subsequently we replace $x=\frac{2\omega}{3}(y-1)$ and treat the transformation of $y' \to x'$ in a similar manner. Thus, the transition probability turns out to be, 
\beq
&&P_{exc}= \frac{g^2\nu^2}{32\pi L^2\omega^2}\bigg|\int_0^{\infty}dxe^{-ix}e^{-x^2/2\omega^2L^2}x^{-2i\nu}\bigg[\frac{5i}{\nu}e^{-2ix\nu/\omega}\nonumber\\
&&+6\, {\rm EI}\left(1+2i\nu,\frac{2i\nu x}{\omega}\right)-6\left(\frac{2i\nu x}{\omega}\right)^{2i\nu}\Gamma(-2i\nu)\bigg]\bigg|^2 .
\label{eq: ei probability}
\eeq
Here, ${\rm EI}$ represents exponential integral, and expanding the function up to the leading order of $x/\omega$ we find \cite{Scully:2017utk}, 
\beq
&&{\rm EI}\left(1+2i\nu,\frac{2i\nu x}{\omega} \right)= \left(\frac{2i\nu x}{\omega}\right)^{2i\nu}\Gamma(-2i\nu)\nonumber\\
&&\exp\left[-4\pi\nu \rm Floor\left(\frac{\pi-Arg(x/\omega)-Arg(i\nu)}{2\pi}\right)\right]+\frac{-i}{2\nu}\nonumber\\
&&-\frac{2\nu x}{\omega(i+2\nu)}~.
\label{error_fu_1}
\eeq
As we are in the very near horizon regime, $x$ is small, and any residual effects in the large $x$ limit gets suppressed by the exponential decay factor. At the same time, as $\omega$ is large, it is legitimate to adopt the $\frac{x}{\omega}<1$ approximation in the following analysis.
Now, the floor function in the above expression can be expressed as, 
\beq
&&\rm Floor\left(\frac{\pi-Arg(x/\omega)-Arg(i\nu)}{2\pi}\right)=  \rm Floor\left(\frac{\pi-0-\pi/2)}{2\pi}\right)\nonumber\\
&&= \rm Floor\left(\frac{\pi/2}{2\pi}\right)=\rm Floor \left(\frac{1}{4}\right) = 0~.
\label{floor_1}
\eeq
Substituting the floor function from Eq. (\ref{floor_1}) into Eq. (\ref{error_fu_1}), one obtains, 
\beq
{\rm EI}\left(1+2i\nu,\frac{2i\nu x}{\omega} \right) = \left(\frac{2i\nu x}{\omega}\right)^{2i\nu}\Gamma(-2i\nu)\nonumber\\
+\frac{-i}{2\nu}-\frac{2\nu x}{\omega(i+2\nu)}~.
\label{error_fl_2}
\eeq
Using Eq. (\ref{error_fl_2}), and implementing $\nu/\omega<<1$, the transition probability as in Eq. (\ref{eq: ei probability}), becomes, 
\beq
&&P_{exc}=\frac{g^2}{32\pi L^2\omega^2} \Bigg|\frac{2L^2\omega^2}{1+2\nu}(i+2\nu) \hg\left(1-i\nu,\frac{3}{2},\frac{-L^2\omega^2}{2}\right)\nonumber\\
&&\Gamma(1-i\nu)+2\sqrt{2}iL\omega\Gamma\left(\frac{1}{2}-i\nu\right)\hg\left(\frac{1}{2}-i\nu, \frac{1}{2},\frac{-L^2\omega^2}{2}\right)\Bigg|^2 ~.\nonumber\\
\label{eq: ffz pexc}    
\eeq
At this stage, we refer our readers to the Appendix (\ref{subsec:probability}) for the detailed derivation of this section. 

\noindent 
Notably, the confluent hypergeometric function has argument $L^2\omega^2/2$. Thus, we study both the argument $L\o/\sqrt{2} >1$ and $L\o/\sqrt{2} <1$. For $L\o/\sqrt{2} >1$, the confluent hypergeometric function can be expanded as follows, 
\beq
\hg(a,b,-z) \approx\Gamma(b)\frac{-z^{-a}}{\Gamma(b-a)}~,
\label{hyper_1}
\eeq
which yields the transition probability as below, 
\beq
P_{exc}= \frac{g^2}{32\pi L^2\omega^2}\Bigg|2\sqrt{\pi}\left[\frac{\Gamma(1-i\nu)}{\Gamma\left(\frac{1}{2}-i\nu\right)}+i\frac{\Gamma\left(\frac{1}{2}-i\nu\right)}{\Gamma(i\nu)}\right]\Bigg|^2~.
\label{trans_finite_2}
\eeq
Further, using trigonometric hyperbolic identities and gamma function identities given below, 
\begin{align}\left.\begin{aligned}
    1.&\quad \Gamma(1-z)\Gamma(z)=\frac{\pi}{\sin(\pi z)}\\
    2.&\quad  |\Gamma(iy)|^2 = \frac{\pi}{y \sinh(\pi y)}\\
 3.&\quad \bigg|\Gamma(1/2+iy)\bigg|^2= \frac{\pi}{\cosh\pi y}
 \end{aligned}\qquad \qquad \quad \qquad \qquad\right\}\label{eq:gamma function identities}
\end{align}
the final form of the transition probability becomes, 
\beq
P_{exc}= \frac{g^2\nu}{2L^2\omega^2}\left(\frac{1}{e^{4\pi\nu}-1}\right)=\frac{g^2\nu r_g c}{2L^2\omega^2}\left(\frac{1}{e^{4\pi\nu r_g/c}-1}\right)~.
\nonumber
\label{trans_finite_3}
\eeq
Following section (\ref{pointlike}), the absorption probability turns out to be, 
\beq
P_{abs}= e^{4\pi\nu r_g/c}P_{exc}~.
\label{abs_exc_finite_1}
\eeq

\noindent
For the other condition $L \o/\sqrt{2}<1$, the confluent hypergeometric function in Eq. (\ref{eq: ffz pexc}), can be expanded as follows, 
\beq
\hg(a,b,z)= \sum_{0}^{\infty} \frac{a^n}{b^n}\frac{z^n}{n!}~.
\label{hyper_2}
\eeq
Using Eq. (\ref{hyper_2}), the transition probability  becomes, 
\beq
&&P_{exc}= \frac{g^2}{32\pi L^2\omega^2}\bigg|2L^2\omega^2\Gamma(1-i\nu)\left(1-\frac{(1-i\nu)L^2\omega^2}{3}\right)\nonumber\\
&&+\frac{2iL\omega}{\sqrt{2}}\Gamma\left(\frac{1}{2}-i\nu\right)\left(1-\left(\frac{1}{2}-i\nu\right)L^2\omega^2\right)\bigg|^2\,.
\label{trans_finite_4}
\eeq
Eq. (\ref{trans_finite_4}) further reduces to, 
\beq
&&P_{exc}=\frac{g^2}{8\pi}\Bigg[\bigg|\nu L\omega\Gamma(-i\nu)\left(1-\frac{(1-i\nu)L^2\omega^2}{3}\right)\bigg|^2+\nonumber\\
&&\bigg|\frac{1}{\sqrt{2}}\Gamma\left(\frac{1}{2}-i\nu\right)\left(1-\left(\frac{1}{2}
-i\nu\right)L^2\omega^2\right)\bigg|^2
 -\frac{\nu L\omega}{\sqrt{2}}\nonumber\\
 &&\bigg\{\Gamma\left(\frac{1}{2}-i\nu\right)\Gamma(i\nu)\left(1-\frac{(1+i\nu)L^2\omega^2}{3}\right)\nonumber\\&&\left(1-\left(1
-2i\nu\right)\frac{L^2\omega^2}{2}\right)+
\Gamma\left(\frac{1}{2}+i\nu\right)\Gamma(-i\nu)\nonumber\\
&&\left(1-\frac{(1-i\nu)L^2\omega^2}{3}\right)\left(1-\left(1
+2i\nu\right)\frac{L^2\omega^2}{2}\right) \bigg\}\Bigg]\,.
\label{trans_finite_5}
\eeq
Writing the Gamma functions in the polar form as below, 
\begin{align}
    \left.\begin{aligned}
        \Gamma(z) &= |\Gamma(z)|e^{i\arg\Gamma(z)}\\
     \quad \arg\Gamma(z)&= - \arg\Gamma(z^*)
    \end{aligned}\right\}\label{eq:gamma arguments}~,
\end{align}
and using Eqs. (\ref{eq:gamma arguments}) and (\ref{eq:gamma function identities}), we have 
\beq
\left\{
\begin{aligned}
\Gamma\left(\frac{1}{2}+i\nu\right) &= \sqrt{\frac{\pi}{\cosh\pi\nu}}e^{-i\arg\Gamma\left(\frac{1}{2}-i\nu\right)} \\
\Gamma\left(\frac{1}{2}-i\nu\right) &= \sqrt{\frac{\pi}{\cosh\pi\nu}}e^{i\arg\Gamma\left(\frac{1}{2}-i\nu\right)} \\
\Gamma(i\nu) &= \sqrt{\frac{\pi}{\nu\sinh\pi\nu}}e^{-i\arg\Gamma(-i\nu)} \\
\Gamma(-i\nu) &= \sqrt{\frac{\pi}{\nu\sinh\pi\nu}}e^{i\arg\Gamma(-i\nu)}
\end{aligned}
\right.
\eeq

\noindent
Thus, the excitation probability turns out to be, 
\beq
&&P_{exc}=\frac{g^2}{8}\bigg[\frac{1}{2\cosh\pi\nu}\left(1-\frac{L^2\omega^2}{2}+i\nu L^2\omega^2\right)^2+\frac{L^2\omega^2\nu}{\sinh\pi\nu}\nonumber\\
&&\left(1-\frac{L^2\omega^2}{3}+\frac{i\nu L^2\omega^2}{3}\right)^2-
\frac{\nu L\omega}{\sqrt{\nu\sinh2\pi\nu}}\bigg\{\bigg(1-\frac{5\omega^2L^2}{6}+\nonumber\\
&&\frac{2i\nu L^2\omega^2}{3}\bigg)e^{i\chi}+\left(1-\frac{5\omega^2L^2}{6}-\frac{2i\nu L^2\omega^2}{3}\right)e^{-i\chi}\bigg\}\bigg]~,
\label{trans_finite_6}
\eeq
where $\chi= \arg\Gamma\left(\frac{1}{2}-i\nu\right)-\arg\Gamma\left(-i\nu\right)$. 
As $L\omega/\sqrt{2}<1$, we keep the terms up to the  $\mathcal{O}\left(L\omega/\sqrt{2}\right)^2$ in the final form of the transition probability which emerges as, 
\beq
&&P_{exc}\approx\frac{g^2}{8}\bigg[\frac{1-L^2\omega^2}{e^{\pi\nu}+e^{-\pi\nu}}+\frac{2\nu L^2\omega^2}{e^{\pi\nu}-e^{-\pi\nu}}-\frac{L\omega\sqrt{2\nu}}{\sqrt{e^{2\pi\nu}-e^{-2\pi\nu}}}\nonumber\\
&&\left\{2\left(1-\frac{5\omega^2L^2}{6}\right)\cos\chi-\frac{4\nu L^2\omega^2}{3}\sin\chi\right\}  \bigg]\,.
\eeq
%
We obtain the absorption probability to be $P_{abs}=P_{exc}$ under the condition $L\o/\sqrt{2}<1$. 

\begin{figure}[h!]
    \centering
    \includegraphics[scale=0.7]{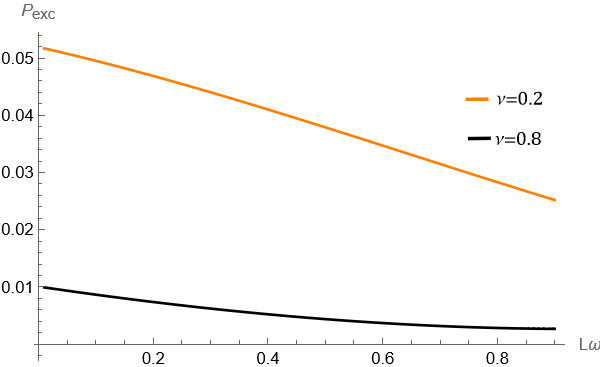}
    \caption{$P_{exc}$ vs $L\o$ plot for finite size detector under the condition $L\o/\sqrt{2}<1$.}
    \label{length_small_2}
\end{figure}

\begin{figure}[h!]
    \centering
    \includegraphics[scale=0.7]{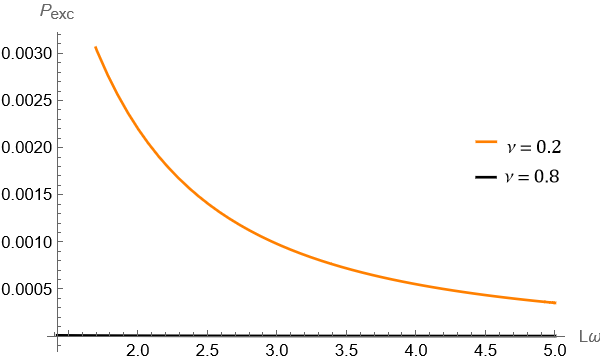}
    \caption{ $P_{exc}$ vs $L\o$ plot for finite size detector under the condition $L\o/\sqrt{2}>1$. }
    \label{length_large_2}
\end{figure}

\noindent
At this stage, we examine how $P_{exc}$ varies with the field frequency  while keeping $L\o/\sqrt{2}$ fixed in both the regimes, $L\o/\sqrt{2}<1$ and $L\o/\sqrt{2}>1$.  
 \begin{figure}[h!]
    \centering
    \includegraphics[scale=0.65]{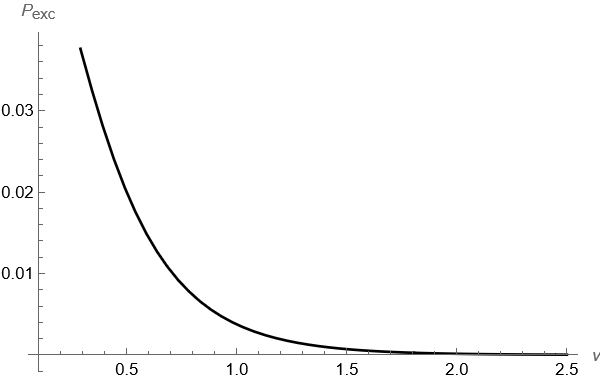}
    \caption{$P_{exc}$ vs $\nu$ plot for finite size detector under the condition $L\o/\sqrt{2}<1$. }
    \label{length_small_1}
\end{figure}
 \begin{figure}[h!]
    \centering
    \includegraphics[scale=0.65]{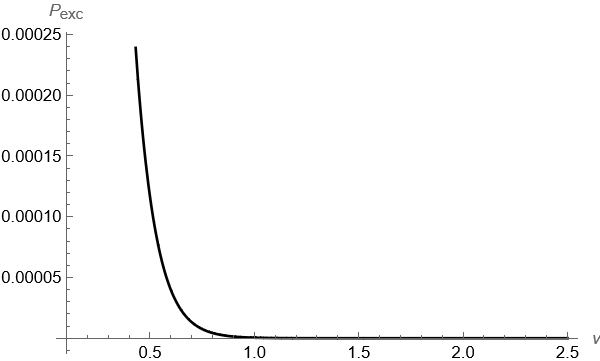}
    \caption{ $P_{exc}$ vs $\nu$ plot for finite size detector under the condition $L\o/\sqrt{2}>1$. }
    \label{length_large_1}
\end{figure}
We identify the enhanced transition probability amplitude in Fig. (\ref{length_small_1}) than that of the Fig. (\ref{length_large_1}). 
Subsequently, in Figures (\ref{length_small_2}) and (\ref{length_large_2}), we plot $P_{exc}$ with respect to $L\o$ considering both the conditions $L\o/\sqrt{2}<1$ and $L\o/\sqrt{2}>1$,  with the constant field frequencies. 
\subsection{Density matrix for the field mode and HBAR entropy: finite size detector case}\label{finite_density}
In this section we find the entropy flux corresponding to the HBAR while the finite size detector is interacting with the momentum of the field. We follow the same procedure as described in the section (\ref{point_density}). 
For the condition $L\o/\sqrt{2}>1$, we find the steady state solution for the density matrix of the field as below
\begin{align}
    \rho_{n,n}^{\mathcal{S}} = e^{-4\pi\nu nr_g/c}(1-e^{-4\pi\nu r_g/c})
    \label{density_finite_1}~.
\end{align}
This suggests that the rate of change of BH entropy due to photon emission in the HBAR process matches the result obtained for a point-like detector in Eq. (\ref{final_entropy_1}). On the other hand, under the condition $L\o/\sqrt{2}<1$, the steady state solution is found to be $\rho^{s,s}=0$. 
This implies that the commonly used approximation for the time rate of change of entropy such as
 \begin{equation}
    \dot{S}_p = - k_B\sum_{n, \nu} \dot{\rho}_{n,n}\,{\rm ln} \rho_{n,n} \implies \dot{S}_p \approx - k_B\sum_{n, \nu} \dot{\rho}_{n,n}\,{\rm ln} \rho^{\mathcal{S}}_{n,n} 
    \nonumber
 \end{equation}
 is no more valid for the case when $\rho^{\mathcal{S}}_{n,n}=0$. 
This hints at the possibility that under the condition $L\o/\sqrt{2}<1$, the system exists in a non equilibrium thermodynamic state, where standard thermal equilibrium and its consequences will no longer hold. 
\section{Examining Wien displacement for point-like and finite size detector}\label{wien_1}
In this section, we present a comparative study by examining the Wien displacement of the HBAR in the context of a point-like detector coupled minimally with the field amplitude \cite{Scully:2017utk} and coupling of the same with the momentum of the field.  
Following the same treatment as in \cite{Das:2023rwg} we write Eq. (\ref{point_exc_6}) as, 
\beq
P(\nu)d\nu= \frac{2 \pi \mathtt{g}^2 G\,\nu M_{\rm BH}}{c^3}\left(\frac{d\nu}{e^{\nu/ T}-1}\right)~,
\label{wein_point_1}
\eeq
where $T=\f{\hbar c^3}{8 \pi G M_{\rm BH}k_B}$, depicts the temperature of the thermal bath due to the HBAR. Using $\hbar=k_B=c=1$, the temperature can be recast as $T=\f{1}{8 \pi G M_{\rm BH}}$. Using $\nu =1/ \lambda$ we write the above equation as follows, 
\beq
P(\lambda)d\lambda= -\f{ 2 \pi \mathtt{g}^2 G\,M_{\rm BH}}{\lambda^3 }\left(\frac{d\lambda}{e^{1/\lambda T}-1}\right)
\label{wein_point_2}
\eeq
Upon varying with respect to the wavelength $(\lambda)$ of the HBAR, the transition probability turns out to be maximum when the denominator of the Eq. (\ref{wein_point_2}) is minimum. 
Thus, writing $z=\lambda^3(e^{1/\lambda T }-1)$, we demand, 
\beq
 \frac{dz}{d\lambda}=0 \,\implies \frac{1}{3\lambda T}= 1-e^{-1/\lambda T}
\label{wein_point_3}
\eeq
At this stage we refer our readers to the section (IV) of \cite{Das:2023rwg} to realise a detailed derivation of the Wein displacement in the context of HBAR phenomenon. Thus, we graphically solve this transcendental equation and obtain $\f{1}{\lambda T}= 2.82$. 
\begin{figure}[h!] 
\centering    \includegraphics[width=0.45\textwidth]{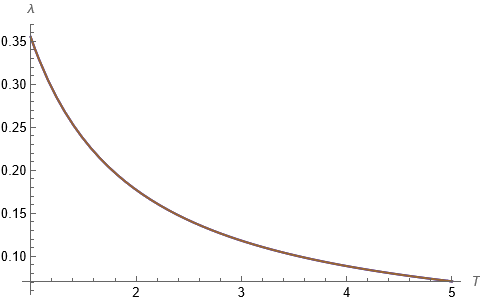} 
\caption{Wein displacement plot with $\lambda$ as a function of $T$. The plots corresponding to the point-like detector (in the present case), the condition $L\omega>\sqrt{2}$ and the minimally coupled model \cite{Scully:2017utk} coincide}
    \label{wein_displacement_1} 
\end{figure}
\noindent
Implementing the same treatment for the finite size detectors for the condition of $L\o/\sqrt{2}> 1$, we obtain the solutions for $\f{1}{\lambda T}= 2.82$.  For a clear depiction, we plot the $\lambda$ vs $T$ in the fig. (\ref{wein_displacement_1}). For $L\o/\sqrt{2} < 1$, we are unable to perform the Wien displacement analysis due to the thermal nature of its excitation probability. 
\section{Discussion}
Particle production in curved/ flat spacetime has been a topic of interest to many physicists since it was first proposed around 1973. Several aspects of this novel mechanism have been explored under the framework of UF effect and its applications, probing the flat/ curved spacetime using UD detector, Hawking radiation from BH, cosmological particle production, and more recently, its role in  quantum entanglement, entanglement harvesting, relativistic quantum information and various new directions constantly emerging.  Over the past few years, techniques from quantum optics have been extensively employed to investigate particle production, and more specifically, acceleration radiation of field quanta. This approach has resulted in several noteworthy developments such as enhancement in the temperature of the thermal bath than that of the standard UF effect, the mechanism of HBAR in curved spacetime, HBAR in the modified gravity theories such as in the context of extra dimensional BH etc. 
In this approach, people generally consider a single mode of the field interacting minimally with the atomic detector, that is, the detector is coupled to the field amplitude. However, literature reveals that minimally coupled atom-field interaction is affected by the IR divergences and its realistic implementation. To the best of our knowledge, this paper presents the first attempt to propose a derivative coupled atom-field interaction for investigating the HBAR phenomenon in curved spacetime. In this work, we explore the HBAR mechanism while examining both point-like and finite size atoms coupled to the momentum of the field. We obtain the transition probability of the system,  where the atom undergoes a transition from its ground state to the higher excited state and simultaneously a one-particle state of the field is generated with a particular frequency.  For the point-like detector, allowing up to the leading order of perturbation theory, we find that the transition probability becomes independent of the frequency of the detector. 
We comment that, physically this can be viewed as the detector coupling to a field which is influenced by the local curvature of spacetime through the $g^{tt}$ component of the metric. We hypothesise that the presence of $g^{tt}$ modifies the sensitivity of the detector to its frequency $\o$, potentially expanding its effective frequency range. This condition may eliminate the sharp resonant condition that we often encounter in the context of standard HBAR process. Therefore, the transition probability becomes independent of $\o$ up to the leading order in perturbation theory, indicating that the background geometry supplies sufficient energy to the atom-field system in order to support the transitions without requiring a sharp resonance. The outcome of this sort is entirely new in the literature of HBAR. However, the modifications in the detector's response function with respect to its frequency in the background of curved spacetime has been reported in literature in the context of UF effect and atom-mirror system \cite{Juarez-Aubry:2014jba, Louko:2007mu, Louko:2014aba}.
Our results for the finite size detector suggest that the transition probability indeed depends on the extent of the detector and based on the conditions, $L\o/\sqrt{2}>1$ or $<1$, the probabilities show distinct behaviour while varying with respect to $\nu$. 
For simplicity and for the sake of discussion, we reduce the above conditions to $L\o\gtrsim 1$ and $L\o\lesssim1$ and 
compare the results for a fixed frequency $\o$. Under the condition $L\o\gtrsim 1$, the detector's length $L$ becomes large compared to the wavelength associated with $\o$. Satisfying the condition that $L$ is also greater than the characteristic wavelength associated with $\nu$, this implies that different parts of the detector interact with the field at different phases which leads to the destructive interference in the response of the detector. Consequently, the overall transition probability decreases.  
Conversely, if $L$ is smaller with respect to the characteristics wavelength of $\o$ in order to comply with the condition $L\o\lesssim1$, it spans approximately a single phase of the field which yields constructive interference in its response. In this case, if $L < \o$, the condition $L$ must be less than the wavelength corresponding to $\nu$ always satisfied. 
We refer our readers to \cite{Lee:2012tvc, Louko:2006zv}, where the modifications in the transition probability of the detector due to its smearing have been reported. The variation of $P_{\rm exc}$ with $L\o$ for fixed field frequency, exhibits similar behaviour as described above. 
Furthermore, for $L\o\lesssim1$, we demonstrate that the steady state solution for the density matrix of the field do not exist. This signifies the breakdown of the approximation used to evaluate the time rate of change of entropy, as employed in \cite{Scully:2017utk}. We comment that under this condition, the system exists in a non equilibrium thermodynamic state, where the usual notions of thermal equilibrium and their consequences no longer hold.

\noindent
The present work captures several new results in the context of HBAR phenomenon in curved spacetime. The interaction of the detector with the momentum of the field plays the central role to dig out these new features. 
In the future, investigating the enhancement of transition probabilities for point-like detectors, verifying the frequency independence of transition probabilities at the next to leading order approximation, and exploring the emergence of non equilibrium conditions for detectors of specific length could be of particular interest. 
\section*{Acknowledgments}
AD acknowledges Soumitra Sengupta and Kinjalk Lochan for many insightful discussions and important suggestions. AK acknowledges the financial support from SRM University, AP.

\section*{Appendix}
\subsection{Detailed derivation of the excitation probability of the finite size detector}\label{subsec:probability}
Following includes the intermediate steps to arrive at Eq. (\ref{eq: ei probability}).   The interaction Hamiltonian for a finite sized detector is 
\begin{align}
    H_{int}&=\hbar g \left(\sigma e^{-i\omega\tau}+\sigma^{\dagger}e^{i\omega\tau}\right)\int dr'F(r') \sqrt{-g}  g^{00} \partial_0\Phi \notag\\
    &= i\hbar g\nu\,\left(\sigma e^{-i\omega\tau}+\sigma^{\dagger}e^{i\omega\tau}\right)\int dr'F(r')\frac{r^3}{r-1}\nonumber\\
    &\quad \qquad \left(a_{\nu}^{\dagger}e^{i\nu(t-r_*)}-a_{\nu}e^{-i\nu(t-r_*)}\right)~,\label{hamil_finite_app}
\end{align}
where $\Phi =\left(a_{\nu}e^{-i \nu t(\t)+i \nu r_{*}(\t)}+{\rm H.C}\right) $. 
Thus the excitation probability becomes, 
\begin{align}
   & P_{exc}= \frac{1}{\hbar^2}\bigg|\int d\tau \bra{1,b}H_{int}\ket{0,a}\bigg|^2\nonumber\\
    &= \frac{\mathtt{g}^2\nu^2}{2\pi L^2}\bigg|\int_1^\infty dr \sqrt{r}e^{-\frac{2i\omega}{3}r^{3/2}}\int_1^{r}dr' \left(\frac{r'^3}{r'-1}\right)e^{-\frac{(r'-1)^2}{2L^2}}\nonumber\\
    & \quad \qquad 
    e^{-i\nu(\frac{2}{3}r'^{3/2,}+r'+2\ln(r'^{1/2}-1)+2r'^{1/2})}\bigg|^2
    \label{pexc_app_1}
    \end{align}
Substituting, $y= r^{3/2}$,
    \begin{align}
     P_{exc}&= \frac{16\mathtt{g}^2\nu^2}{162\pi L^2}\bigg|\int_1^{\infty}dy e^{-\frac{2i\omega}{3}y} \int_1^y dy' 
   \left(\frac{y'^{5/3}}{y'^{2/3}-1}\right)\nonumber\\
   &\quad e^{-\frac{(y'^{2/3}-1)^2}{2L^2}}e^{-i\nu(\frac{2}{3}y'+y'^{2/3}+2y'^{1/3}+2\ln(y'^{1/3}-1))}\bigg|^2 
   \label{pexc_app_2}
   \end{align}
   Further, replacing $x=\frac{2\omega}{3}(y-1)$,
   \begin{align}
     P_{exc}&= \frac{\mathtt{g}^2\nu^2}{2\pi L^2\omega^4}\bigg|\int_0^{\infty}dx\,e^{-ix}\int_0^{x}dx' \frac{\left(\frac{3x'}{2\omega}+1\right)^{5/3}}{f(x')}\nonumber\\
     &\quad   e^{\frac{-f(x')^2}{2L^2}}e^{-i\nu\,\Omega(x')}\bigg|^2\label{pexc_app_3}
\end{align}
where,
\begin{align*}
&\Omega(x') =\frac{2}{3}\bigg(\frac{3x'}{2\omega}+1\bigg)+2\bigg(\frac{3x'}{2\omega}+1\bigg)^{1/3}\\
&\quad\quad +\bigg(\frac{3x'}{2\omega}+1\bigg)^{2/3}+2\ln\bigg(\left(\frac{3x'}{2\omega}+1\right)^{1/3}-1\bigg) \\
&= \frac{2}{3}\bigg(\frac{3x'}{2\omega}+1\bigg)+2\bigg(1+\frac{1}{3}\bigg(\frac{3x'}{2\omega}\bigg)+..\bigg)+ \\
& \quad\bigg(1+\frac{2}{3}\bigg(\frac{3x'}{2\omega}\bigg)
+... \bigg)+2\ln\bigg(1+\frac{1}{3}\bigg(\frac{3x'}{2\omega}\bigg)+..-1\bigg)
\end{align*}
Employing the condition $1/\o \ll 1$, we obtain, 
\begin{align}
&\Omega(x') \approx \frac{2x'}{\omega}+3+2\ln\left(\frac{x'}{2\omega}
\right) \label{omega_app_2}
\end{align}
and
\begin{align}
f(x')= \bigg(\frac{3x'}{2\omega}+1\bigg)^{2/3}-1 \approx \frac{x'}{\omega}
\label{f_x_app_1}
\end{align}
Using these approximations, the excitation probability turns out to be, 
\begin{align}
&P_{exc}\approx  \frac{\mathtt{g}^2\nu^2}{2\pi L^2\omega^4}\bigg|\int_{0}^{\infty}dx\,e^{-ix} \int_{0}^{x}dx'\frac{\omega}{x'}\left(1+\frac{5x'}{2\omega}\right)\nonumber\\
&\quad \qquad e^{-\frac{x'^2}{2\omega^2L^2}}e^{-2i\nu x'/\omega}x^{-2i\nu}\bigg|^2\nonumber \\
&\approx \frac{\mathtt{g}^2\nu^2}{2\pi L^2\omega^4}\bigg|\int_{0}^{\infty}dx\,e^{-ix}e^{-\frac{x^2}{2\omega^2L^2}}\nonumber\\
&\quad \qquad \int_{0}^{x}dx'\frac{\omega}{x'}\left(1+\frac{5x'}{2\omega}\right)e^{-2i\nu x'/\omega}x^{-2i\nu}\bigg|^2 
\label{p_exc_app_4}
\end{align}
For the above equation, the gaussian term inside $\int dx'$ varies slowly compared to the other terms. And hence, it is taken out of the integral with $x'\rightarrow x$.
\begin{align}
     &P_{exc}= \frac{\mathtt{g}^2\nu^2}{32\pi L^2\omega^2}\Bigg|\int_0^{\infty}dxe^{-ix}e^{-x^2/2\omega^2L^2}x^{-2i\nu}\notag \\
     &\bigg[\frac{5i}{\nu}e^{-2ix\nu/\omega}+6 {\rm EI}\left(1+2i\nu,\frac{2i\nu x}{\omega} \right)-6\left(\frac{2i\nu x}{\omega}\right)^{2i\nu}\notag \\
     &\Gamma(-2i\nu)\bigg]\Bigg|^2 
     \label{p_exc_app_5}
\end{align}
\subsection{Density Matrix Formulation}\label{subsec:density matrix}
Following the procedure as mentioned in \cite{Azizi:2021qcu}, the rate of change of the density matrix is given as,  (averaged over a distribution of random injection times)
\begin{align}
    \dot{\rho}^P = -{\k}\int_{\tau_i}^{\tau_f}\int_{\tau_i}^{\tau ^{\prime}}d\tau^{\prime}d\tau^{\prime\prime}\tra\left[ H(\tau^{\prime}),\notag \left[ H(\tau^{\prime\prime}),\rho^ {AP}(\tau_i)\right]\right]~.\label{eq: denmat1}\\
\end{align}
Here $\k$, is the injection rate of the atom with $\k = \Delta \mathcal{N}/\Delta t$.
$A$ denotes the atom and the index $P$ stands for the field. 
Tracing over the atom's degrees of freedom one gets, 
\begin{align}
&\text{Tr}_A \Big[ H(\tau'), [H(\tau''), \rho^{AP}(\tau_i)] \Big]\notag\\
&= \text{Tr}_A \Big[ H' [H'', \rho^{AP}(\tau_i)] 
- [H'', \rho^{AP}(\tau_i)] H' \Big]
\label{trace_app_1}
\end{align}
For a point-like detector, we write the Hamiltonian as below, 
\begin{align}
H(\tau)=\mathtt{g}\, \mu(\tau) \pi(\tau)
= \mathtt{g}\, \frac{r^3}{r - 1} \mu(\tau) \partial_0 \Phi(\tau)
\label{hamil_app_2}
\end{align}
We obtain $\partial_0\Phi(\tau) = i\nu\left( a_\nu^{\dagger} e^{i\nu(t-r_*)}-{\rm H.C}\right) = -i\nu\varphi(\tau)$, where $\varphi(\tau) = \left( a_\nu^{\dagger} e^{i\nu(t-r_*)}-{\rm H.C}\right)=\left(a_\nu^{\dagger}\phi-{\rm H.C}\right)$. Here $\phi=e^{i \nu (t-r_{*})}$ [see sec. (\ref{trajectory_atom})]. 
Therefore the Hamiltonian becomes, 
   \begin{equation}
       H(\tau) = ig\nu\frac{r^3}{r-1}\mu(\tau)\varphi(\tau) =ig\nu f(r)\mu(\tau)\varphi(\tau)
   \end{equation}
   with $f(r)=\frac{r^3}{r-1}$.
With this, the trace is, 
\beq
&&\tra\bigg[H'[H'',\rho^{AP}(\tau_i)]-[H'',\rho^{AP}(\tau_i)]H\bigg]\nonumber\\
&&= -g^2\nu^2\tra \bigg[f^2(r)\mu'\varphi'\mu''\varphi''\rho^{AP}_i-f^2(r)\mu''\varphi''\rho^{AP}_i\mu'\varphi' \nonumber\\
&&+f^2(r)\rho^{AP}_i\mu''\varphi''\mu'\varphi'-f^2(r)\mu'\varphi'\rho^{AP}_i\mu''\varphi''\bigg]\nonumber\\
&&= -g^2\nu^2\bigg[f^2(r)\varphi'\varphi''\rho^P_i\tra(\mu'\mu''\rho_i^A)+\nonumber\\
&&f^2(r)\rho_i^P\varphi''\varphi'\tra(\rho_i^A\mu''\mu')-f^2(r)\varphi'\rho_i^P\varphi''\tra(\mu'\rho_i^A\mu'')-\nonumber\\
&&f^2(r)\varphi''\rho_i^P\varphi'\tra(\mu''\rho_i^A\mu')\bigg]\nonumber\\
&&= -g^2\nu^2f^2(r)\bigg[\varphi'\varphi''\rho^P_ie^{-i\omega\tau'}e^{i\omega\tau''}+\rho_i^P\varphi''\varphi'e^{i\omega\tau'}e^{-i\omega\tau''}\nonumber\\
&&-\varphi'\rho_i^P\varphi''e^{i\omega\tau'}e^{-i\omega\tau''} -\varphi''\rho_i^P\varphi'e^{-i\omega\tau'}e^{i\omega\tau''}\bigg]\label{trace_app_0}
\eeq,
where in the last step, we trace over the atom's degrees of freedom $(\ket{a},\ket{b})$ and take the initial state of the atom's density matrix to be $\ket{a}\bra{a}$. Since, $\tau'>\tau''$, the rate of change in the density matrix can be written as, 
\begin{align}
&\dot{\rho}^P = g^2 \nu^2 \kappa \Bigg[
    \int_{\tau^{\prime}>\tau^{\prime\prime}} d^2\tau\, f^2(r) \varphi^{\prime} \varphi^{\prime\prime} \rho^P e^{-i\omega\tau'} e^{i\omega\tau''} \nonumber\\
&+ \int_{\tau^{\prime}>\tau^{\prime\prime}} d^2\tau\, f^2(r) \rho^P \varphi^{\prime\prime} \varphi^{\prime} e^{i\omega\tau'} e^{-i\omega\tau''}- \int_{\tau^{\prime}>\tau^{\prime\prime}} d^2\tau f^2(r) \nonumber\\
&\varphi^{\prime} \rho^P \varphi^{\prime\prime} e^{i\omega\tau'} e^{-i\omega\tau''} - \int_{\tau^{\prime}>\tau^{\prime\prime}} d^2\tau\, f^2(r) \varphi^{\prime\prime} \rho^P \varphi^{\prime} e^{-i\omega\tau'} e^{i\omega\tau''}
\Bigg]~.
\label{density_app_1}
\end{align}
 where $d\tau'd\tau''=d^2\tau$. 
We can symmetrize this entire integration by swapping the region of integration of the second and the fourth terms of the above equation, that is, swapping $\tau^{\prime} \leftrightarrow \tau^{\prime\prime}$ and $\varphi^{\prime} \leftrightarrow \varphi^{\prime\prime}$. By naming the region 
$\tau^{\prime}>\tau^{\prime\prime}$ as region $I$ and $\tau^{\prime\prime}>\tau^{\prime}$ as region $II$, this yields, 
\begin{align}
&&\dot{\rho^P}=g^2\nu^2\k \Bigg[\bigg(\int_{I}d^2\tau f^2(r)\varphi^{\prime}\varphi^{\prime\prime}\rho^P+\int_{II}d^2\tau f^2(r)\rho^P\nonumber\\
&&\varphi^{\prime}\varphi^{\prime\prime}\bigg)e^{-i\omega\tau'}e^{i\omega\tau''}- \int_{I+II}d^2\tau 
f^2(r)\varphi^{\prime}\rho^P\varphi^{\prime\prime}e^{i\omega\tau'}e^{-i\omega\tau''}\Bigg]
\label{eq: after swapping}
\end{align}
For brevity, we drop the superscript $P$ from here on. We now find the elements of the reduced density matrix and use the notation $\rho_{nm}=\braket{n|\rho|m}$. We have three elements to be computed from the above expression: $\bra{n}\varphi^{\prime}\varphi^{\prime\prime}\rho\ket{n} ,\quad \bra{n} \rho\varphi^{\prime}\varphi^{\prime\prime}\ket{n},\quad \bra{n}\varphi^{\prime}\rho\varphi^{\prime\prime}\ket{n} $ . 

\onecolumngrid
\hrulefill
\begin{align}
        \bra{n}\varphi^{\prime}\varphi^{\prime\prime}\rho\ket{n} &= \bra{n}\left(a_{\nu}^{\dagger\prime}\phi^{\prime}-a_{\nu}^{\prime}\phi^{\prime}\right)\left(a_{\nu}^{\dagger\prime\prime} \phi^{\prime\prime}-a_{\nu}^{\prime\prime}\phi^{*\prime\prime}\right)\rho\ket{n}\notag\\
        &= \bra{n}\left(a_{\nu}^{\dagger\prime}\phi^{\prime}a_{\nu}^{\dagger\prime\prime} \phi^{\prime\prime}-a_{\nu}^{\dagger\prime}\phi^{\prime}a_{\nu}^{\prime\prime}\phi^{*\prime\prime} - a_{\nu}^{\prime}\phi^{*\prime}a_{\nu}^{\dagger\prime\prime}\phi^{\prime\prime}+a_{\nu}^{\prime}\phi^{*\prime}a_{\nu}^{\prime\prime}\phi^{*\prime\prime}\right)\rho\ket{n}\notag\\
        &= \phi^{\prime}\phi^{\prime\prime}\bra{n}a_{\nu}^{\dagger\prime}a_{\nu}^{\dagger\prime}\rho\ket{n}-\phi^{\prime}\phi^{*\prime\prime}\bra{n}a_{\nu}^{\dagger\prime}a_{\nu}^{\prime\prime}\rho\ket{n}-\phi^{*\prime}\phi^{\prime\prime}\bra{n}a_{\nu}^{\prime}a_{\nu}^{\dagger\prime\prime}\rho\ket{n}+\phi^{*\prime}\phi^{*\prime\prime}\bra{n}a_{\nu}^{\prime}a_{\nu}^{\prime\prime}\rho\ket{n}\notag\\
        &= \sqrt{n(n-1)}\phi^{\prime}\phi^{\prime\prime}\bra{n-2}\rho\ket{n}-n\phi^{\prime}\phi^{*\prime\prime}\bra{n}\rho\ket{n}-\notag\\        &\qquad\qquad(n+1)\phi^{*\prime}\phi^{\prime\prime}\bra{n}\rho\ket{n}+ \sqrt{(n+1)(n+2)}\phi^{*\prime}\phi^{*\prime\prime}\bra{n+2}\rho\ket{n}\notag\\
        &= \sqrt{n(n-1)}\phi^{\prime}\phi^{\prime\prime}\rho_{n-2,n}-n\phi^{\prime}\phi^{*\prime\prime}\rho_{n,n}-(n+1)\phi^{*\prime}\phi^{\prime\prime}\rho_{n,n}+ \sqrt{(n+1)(n+2)}\phi^{*\prime}\phi^{*\prime\prime}\rho_{n+2,n}\label{eq:dens_matrix_el_1}     
        \end{align}
        Similarly, we get
        \begin{align}  
        \bra{n} \rho\varphi^{\prime}\varphi^{\prime\prime}\ket{n} &=  \sqrt{(n+1)(n+2)}\phi^{\prime}\phi^{\prime\prime}\rho_{n,n+2} - n\phi^{\prime}\phi^{*\prime\prime}\rho_{n,n}-(n+1)\phi^{*\prime}\phi^{\prime\prime}\rho_{n,n}+\sqrt{n(n-1)}\phi^{*\prime}\phi^{*\prime\prime}\rho_{n,n-2}\label{eq:dens_matrix_el_2}\\ 
         \notag\\
        \bra{n}\varphi^{\prime}\rho\varphi^{\prime\prime}\ket{n}&= \sqrt{n(n+1)}\phi^{\prime}\phi^{\prime\prime}\rho_{n-1,n+1}-n\phi^{\prime}\phi^{*\prime\prime}\rho_{n-1,n-1}-(n+1)\phi^{*\prime}\phi^{\prime\prime}\rho_{n+1,n+1}+\sqrt{n(n+1)}\phi^{*\prime}\phi^{\prime\prime}\rho_{n+1,n-1}\label{eq:dens_matrix_el_3}
    \end{align}
As the injection time is random, we consider only the diagonal elements of the matrix where for $\rho_{m,n}, \,\, m=n$. The first two terms of (\ref{eq: after swapping}) 
\begin{align}
    &\braket{n|\left(\displaystyle\int_{I}d^2\tau f^2(r)\varphi^{\prime}\varphi^{\prime\prime}\rho^P+\int_{II}d^2\tau f^2(r)\rho^P\varphi^{\prime}\varphi^{\prime\prime}\right)e^{-i\omega\tau'}e^{i\omega\tau''}|n}\nonumber\\\
    &= e^{-i\omega\tau'}e^{i\omega\tau''}\bigg(-\displaystyle \int_{I+II}d^2 \tau f^2(r)\,n\phi^{\prime}\phi^{*\prime\prime}\rho_{n,n} \,+\, (n+1)\phi^{*\prime}\phi^{\prime\prime}\rho_{n,n}\bigg)\nonumber\\
    &= -\bigg(\displaystyle\int_{\tau_i}^{\tau_f}d\tau'f(r)\phi^{\prime}e^{-i\omega\tau'}\int_{\tau_i}^{\tau_f}d\tau''f(r)\phi^{*\prime\prime}e^{i\omega\tau''}\bigg)n\rho_{n,n}-\bigg(\int_{\tau_i}^{\tau_f}d\tau'f(r)\phi^{*\prime}e^{-i\omega\tau'}\int_{\tau_i}^{\tau_f}d\tau''f(r)\phi^{\prime\prime}e^{i\omega\tau''}\bigg)(n+1)n\rho_{n,n}\nonumber\\
    &= -|I_{a,s}|^2n\rho_{n,n}-|I_{e,s}|^2(n+1)n\rho_{n,n}~,\label{matrix_app_1}
\end{align}
where $|I_{e,s}|^2 = \bigg|\displaystyle\int d\tau f(r) \psi e^{i\omega\tau}\bigg|^2$ and $|I_{a,s}|^2 = \bigg|\displaystyle\int d\tau f(r)\psi e^{-i\omega\tau}\bigg|^2$. 
Similarly, we get third and the fourth terms as, 
\begin{align}
    &\braket{n|\int_{I+II}d^2\tau f^2(r)\varphi^{\prime}\rho^P\varphi^{\prime\prime}e^{i\omega\tau'}e^{-i\omega\tau''}|n}= \int_{I+II}d^2\tau f^2(r)(-n\phi^{\prime}\phi^{*\prime\prime}\rho_{n-1,n-1}-(n+1)\phi^{*\prime}\phi^{\prime\prime}\rho_{n+1,n+1})e^{i\omega\tau'}e^{-i\omega\tau''}\notag\\
    &= -|I_{e,s}|^2n\rho_{n-1,n-1}-|I_{a,s}|^2(n+1)\rho_{n+1,n+1}\label{eq:matrix_app_2}
\end{align}
\hrulefill
\twocolumngrid
Therefore the reduced density matrix is now,
\begin{align}
    \dot{\rho^P}&=g^2\nu^2\k\bigg[-|I_{a,s}|^2n\rho_{n,n}-|I_{e,s}|^2(n+1)n\rho_{n,n}\notag\\
    &\quad\qquad  +|I_{e,s}|^2n\rho_{n-1,n-1}+|I_{a,s}|^2(n+1)\rho_{n+1,n+1}\bigg]\notag\\
    &= -\k P_{exc}((n+1)\rho_{n,n}-n\rho_{n-1,n-1})\notag\\
    &\quad \qquad -\k P_{abs}(n\rho_{n,n}-(n+1)\rho_{n+1,n+1})\notag\\
    &= -R_{e}((n+1)\rho_{n,n}-n\rho_{n-1,n-1})\notag\\
    &\quad \qquad -R_{a}(n\rho_{n,n}-(n+1)\rho_{n+1,n+1})\label{eq:density matrix appendix}
\end{align}
where $P_{exc/abs}=g^2\nu^2|I_{e,s/a,s}|^2$, and $R_{e}$ and $R_{a}$ are the emission rate and the absorption rate respectively, with $R_{e/a}=\k P_{exc/abs}$. The steady state solution of the density matrix will thus be, 
\beq
    \rho_{n,n}^{\mathcal{S}}= C\left(\frac{R_{e}}{R_{a}}\right)^n \label{eq:density matrix steady state}
\eeq
where $C$ is a normalization constant that can be evaluated using the Trace property of a density matrix.
\beq
   {\rm Tr}\rho_{n,n}^{\mathcal{S}}&= \sum_{n=0}^{\infty}\rho_{n,n}^{\mathcal{S}}=C\sum_{n=0}^{\infty}\left(\frac{R_e}{R_a}\right)^n =1 \label{eq:trace}
\eeq
For a point like detector,
\begin{align}
    \rho_{n,n}^{\mathcal{S}}&= C\left(\frac{R_e}{R_a}\right)^n=C\left(\frac{P_{exc}}{P_{abs}}\right)^n =Ce^{-4\pi\nu n r_g/c}\notag\\
     {\rm Tr}\rho_{n,n}^{\mathcal{S}}&= \sum_{n=0}^{\infty}\rho_{n,n}^{\mathcal{S}}\notag\\
     &=C\sum_{n=0}^{\infty}e^{-4\pi\nu n r_g/c} =C\left(\frac{1}{1-e^{-4\pi\nu r_g/c}}\right)=1\notag\\
     \therefore  \rho_{n,n}^{\mathcal{S}}&= (1-e^{-4\pi\nu r_g/c})e^{-4\pi\nu n r_g/c}\label{eq:final_solution_steady_state}
\end{align}
For the finite-sized detector, with the Hamiltonian $ H_{int}= g \mu(\tau)\int dr'F(r') \sqrt{-g}  g^{00} \partial_0\phi$, we get the same form for steady state solution of the reduced density matrix as (\ref{eq:density matrix steady state}) when we follow the same steps.  For the  case (where $L\omega>\sqrt{2}$),   we get
\begin{align}
    P_{abs}&= e^{-4\pi\nu rg/c}P_{exc}\label{eq:pabs_pexc_finite_size}
\end{align}
Hence, the steady state solution of density matrix is
\beq
     \rho_{n,n}^{\mathcal{S}}&= C\left(\frac{R_e}{R_a}\right)^n = (1-e^{-4\pi\nu r_g/c})e^{-4\pi\nu n r_g/c}\label{eq:density_matrix_finite_size}
\eeq
which is same as that of the point-like detector. However, in case of $L\omega<\sqrt{2}$, $P_{e,s}=P_{a,s}$, and thus $C=0$ implying $\rho_{n,n}^{\mathcal{S}}=0   $

\end{document}